\documentclass{mpe_report}

\usepackage{psfig,graphicx,epsfig}
\usepackage{color}
\usepackage{amsmath,amssymb,epic,eepic,array}

\begin{document}

\pagenumbering{arabic}
\setcounter{page}{185}

\renewcommand{\FirstPageOfPaper }{185}\renewcommand{\LastPageOfPaper }{188}\def\beq{\begin{equation}}
\def\eneq{\end{equation}}
\def\simgt{\lower.5ex\hbox{$\; \buildrel > \over \sim \;$}}
\def\simlt{\lower.5ex\hbox{$\; \buildrel < \over \sim \;$}}

\title{Neutron Stars in  Microquasars}
\author{M. Massi}  
\institute{Max--Planck--Institut f\"ur 
Radioastronomie,
Auf dem H\"ugel 69, D-53121 Bonn, Germany}
\maketitle

\begin{abstract}
We discuss the  ``basic" condition for an accreting
neutron star to become a microquasar, i.e. ejecting
relativistic particles orthogonal to the accretion disk instead of
confining disk-material down to the magnetic poles and creating
the two emitting caps typical for a X-ray pulsar.
Jet creation is prevented for B$\ge 10^{12}$ G  independent of the accretion
rate.
This excludes the possibility for a classic X-ray pulsar to develop
a "microquasar-phase" and is consistent with  the lack of radio emission  from such pulsar
systems.
Millisecond accretion-powered pulsars, on the contrary, may show a "microquasar-
phase",
where $B <  10^{7.5} $G is valid,
whereas the limit for Z sources is
$B< 10^{8.2} $G.
The implication of our analysis is that
the jet might be the suitable agent of angular momentum sink for
millisecond accretion-powered pulsars.
\end{abstract}

\section{Introduction}

X-ray binaries are  stellar systems formed by two stars
of a very different nature:  A normal star (acting as a mass donor)
 and a compact object (the accretor) that either can
 be a neutron star or a black hole.
However, the classification of the  X-ray binaries (XRBs)    into
 Low Mass X-ray Binary  and High Mass X-ray Binary systems
leaves unspecified  the nature of the accreting
object and exclusively is based on 
the companion star
(van Paradijs \& McClintock 1996),
for HMXRB  young bright stars (O-B) or for LMXRB old stars (later than G) .
An accreting object (as result of a supernova explosion) close to a normal star
is an extremely rare binary among stellar systems and in fact 
up to now   only 280 X-ray binary systems of this type are known
(Liu et al. 2000, 2001). The microquasars (Massi 2005) 
are the 17 XRB systems, where  
 high resolution radio interferometric techniques
have  shown  the presence of collimated   jets or 
  a flat spectrum has been observed (indirect evidence for an expanding  continuous jet, Fender 2004).

\section{Magnetohydrodynamic jet production}
 Numerical simulations show that the launch of jets involves a weak 
large-scale poloidal magnetic field anchored
 in rapidly 
rotating disks/compact objects  (Meier et al. 2001).
The geometry of this field is analogous to that of solar coronal holes
and generated by the dynamo process (Blackman \& Tan 2004).

The strength of the 
large-scale poloidal field must be low enough,
that the  plasma pressure ($P_p$)dominates the magnetic field pressure ($P_B$). 
Only under that  condition, $P_B < P_p$, the differentially rotating  disk is able to bend
the magnetic field lines, resulting in a magnetic spiral
(Meier et al. 2001).
Because of  the increasing  compression of the magnetic field lines
the magnetic pressure will grow 
and may  become larger than the gas pressure at
the surface of the accretion disk, where the density is lower.
There, the magnetic field  becomes "active" (i.e. dynamically
dominant) and  the plasma has to follow the twisted magnetic field lines, 
creating  two spinning plasma-flows.
As recently proved for the bipolar outflows from young stellar objects,
these rotating plasma-flows take angular momentum 
away from the  disk (magnetic braking):
The angular momentum transport rate of the jet can be two thirds or more of the estimated rate transported through the relevant portion of the disk (Woitas et al. 2005).
This loss of  angular momentum 
slows down the disk material to sub-Keplerian rotation and therefore the
disk matter can finally accrete onto the central object.
The accreting  material further pulls
the deformed magnetic field with it in that way increasing the magnetic field
compression 
and   magnetic reconnection may occur (Matsumoto et al 1996).
The stored magnetic energy is released and the field returns to the
state of minimum energy (i.e. untwisted).

\section{The Alfven radius}
\label{alfven}
As seen in the previous section, the ``basic" condition for ejection
is a weak magnetic field. But how weak? 
In case of neutron stars we can take observed values, which
allow a quantitative estimate.
In this  case  there is an additional
stellar magnetic field together with the disk-field.
Let us assume that the stellar field dominates.
The stellar  field  will be bent in the sweeping spiral
(discussed above) only 
if  the  magnetic pressure $B^2 \over {8 \pi} $ is less than
the hydrodynamic pressure $\rho v^2$ of the accreting material.
We here show that the existence of the above conditions
can easily be verified by plotting  the  
Alfv\'en radius (normalised to the stellar surface) vs accretion rate
and magnetic field strength.
The Alfv\'en radius is the
radius at which the magnetic and plasma pressure balance each other.
If $R_A/R_*< 1$, the  field will be twisted ($R_*$ is the stellar radius).

The Alfv\'en radius depends  on the strength, 
 on the (bipolar or multipolar) topology
of the magnetic field and on the mode of accretion 
(spherically symmetric  or disk-like).
In case of a spherically symmetric accretion, 
 the mass accretion rate $\dot{M}$ is equal to 
$4 \pi R^2 \rho v^2$ (Longair 1994), where $v$ is the infall velocity 
$v=(2 G M_* / R) ^{1/2}$.
For a magnetic dipole field with a surface magnetic field  $B_*$,
${B/ B_*}=[{R_* / R}]^{3}$.
Therefore the parameter $R_A/R_*$  in terms of accretion rate $\dot{M}$ 
and surface magnetic field $B_*$
is equal to: 
\beq
R_A/R_* \simeq  0.87 \left ( { B_*\over 10^8} \right)^{4/7} \left ( {\dot M\over 10^{-8}} \right)^{-2/7}.
\label{ra}
\eneq
The equation has been calculated for a neutron star with 
a mass and a radius of respectively $M_*=1.44~ \rm{M}_{\odot}$ and $R_*= 9$ km 
(Titarchuk $\&$ Shaposhnikov  2002).

\section{Neutron star X-ray binaries}

Table 1 shows the  ranges, available in the literature, for accretion rate and magnetic field strength of neutron stars in X-ray binary systems.
Including the classic accretion-powered pulsars
the  interval for $B$  ranges over more than 4 orders of magnitude:
From classic accretion-powered pulsars
with fields above 10$^{12}$G, 
to the low  value of 
10$^{7-8}$G of  millisec-pulsars and atoll sources. 
The interval for  accretion rate covers several orders of
magnitude, too, 
 from less than 0.1\%  Eddington critical rate (1.6 $10^{-8} \rm{M}_{\odot}~\rm{yr}^{-1})$ 
for  millisecond-pulsars to  Eddington critical rate 
for the  Z sources (see references in Table 1).

\begin{figure}[]
\centerline{\psfig{file=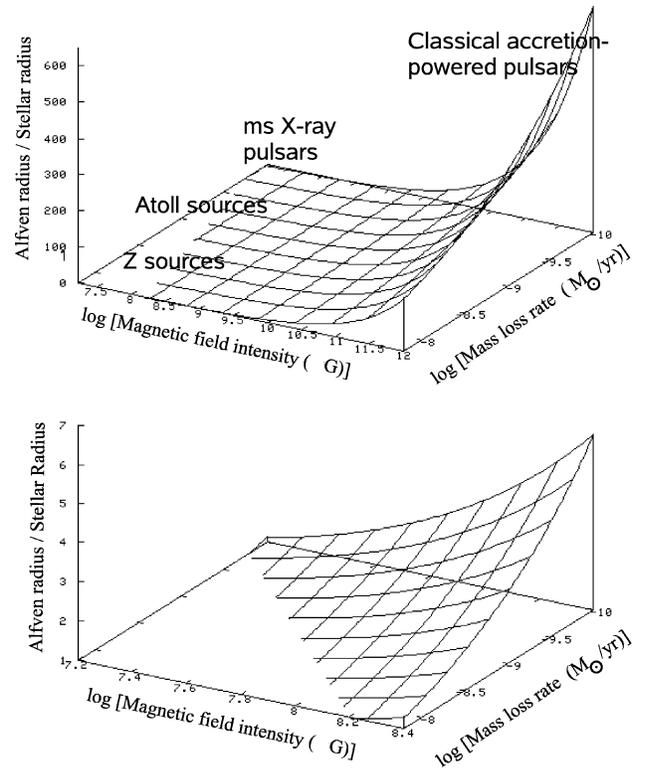,width=8.8cm,angle=0,clip=} }
\caption{ 3-D plot of the Alfven radius normalised to the stellar radius. 
The white area  indicates the region with
 $R/R_*\leq$1 , where 
the basic condition for ejection to occur is fullfilled, i.e. plasma pressure dominates the magnetic field
pressure.  This region is more clearly shown in the bottom of the figure. 
}
\label{3-D}
\end{figure}

In Fig.~\ref{3-D} we show a 3-D plot of the parameter $R_A/R_*$ as function
of both,  accretion rate 
and magnetic field strength.
The value $R_A/R_*$ is plotted only above  unity, i.e. 
the "white" area refers to values of $R_A/R_*<1$. This is the region,
where accretion rate and magnetic field strength are combined in such a way
that the
stellar field is nowhere dynamically important.
This means the white region of Fig.~\ref{3-D}  is the part of the parameter space 
 where potential 
microquasars can exist.   
One can see  in Fig. ~\ref{3-D} that this region is rather  small for the
given large range of $B$ and $\dot M$. 

\begin{table}
\begin{center}
\caption[]{Neutron Stars: Accretion Rate and Magnetic Field Strength}
\begin{tabular}{lll}
\hline \hline \noalign{\smallskip}
class& ${\dot M\over 10^{-8}}  (\rm{M}_{\odot}~\rm{yr}^{-1})$& $B_*$ (G)  \\
\noalign{\smallskip} \hline \noalign{\smallskip}
ms X-ray &  0.001 (a),0.01 (b, c)& $ 3\times10^7$(b,c),$3 \times 10^8$ (b) \\
& 0.03(a),0.07(c)&  \\
\noalign{\smallskip} \hline \noalign{\smallskip}
Atoll   & 0.01-0.9 (d)  & $ 3\times 10^7$ (e),$ 10^8$    (f)  \\
\noalign{\smallskip} \hline \noalign{\smallskip}
Z  sources& $0.5 {\rm (d),} \ge$1 (d)     &$10^{7-8}$ (g), $3\times 10^8$ (e), $10^9$ (f)       \\
\noalign{\smallskip} \hline
\end{tabular}
\end{center}
\label{table:mdot}
[a] Chakrabati \& Morgan 1998  
[b] Lamb $\&$ Yu 2005 
[c] Gilfanov et al. 1998 
[d] van der Klis 1996 
[e] Zhang \& Kojima 2006 
[f] van der Klis 1994
[g] Titarchuk et al. 2001 
\end{table}

\subsection{Classical accretion-powered pulsars}\label{classic}
Classical accretion-powered pulsars have periods of the order of one second or more. In this sense they are also called "slow" accretion-powered pulsars in comparison
to the milliseconds-pulsars (Sect.~\ref{millisec}).
Only five classic accretion-powered pulsars
have been found in LMXRBs, whereas the vast majority are found in HMXRB systems
(Psaltis 2004).

As one derives from Fig.~\ref{3-D}, classic accretion-powered pulsars with  magnetic fields of $10^{12}$G
have  $R_A/R_*>>1$  
even for  accretion rates comparable to the Eddington critical rate.
The stellar field therefore is dynamically dominant. 
In this case the plasma is forced to
move along the magnetic field lines,
converges  onto the magnetic poles of the neutron star and there releases
its energy creating
two X-ray emitting caps. In the case of a misalignment between
the rotation and the  magnetic axis,  X-ray pulses are produced (Psaltis 2004).

 The value $R_A/R_*>>1$, therefore,
{\bf excludes} the formation of jets in accreting pulsars for any accretion rate.
Our  result agrees  with the observations.
A deep search for radio emission from X-ray pulsars has been performed
by Fender et al. (1997) and none of the pulsar candidates has been detected.
The lack of radio emission is statistically discussed
by Fender et al. (1997) and they conclude
 that X-ray pulsations and radio emission
from X-ray binaries are strongly anti-correlated.
More recent observations (Fender $\&$ Hendry 2000; Migliari $\&$ Fender 2005)
 again confirm that  
none of the high-magnetic-field X-ray pulsars is
 a source of synchrotron radio emission.

\subsection{Atoll and Z-sources}

The LMXRBs with neutron stars have been divided into two subclasses called  Atoll-type
 (the largest class) and Z-type, based on their timing and spectral properties
 observed by Hasinger $\&$ van der Klis (1989).
The differences reflect different values of accretion rates, but also
different values for the magnetic field strength (see Table 1).
For each of the two types, Atoll-sources and  Z-sources, the accretion rate  
is different in the different spectral states:  For the Atoll-sources there are two states
(island and banana) and for the Z-type sources there are three ones 
(horizontal-, flaring-, and normal-branch).

Two Z sources (Circinus X-1 and Scorpius X-1)
are microquasars.   Therefore, the condition  
$R_A/R_*\leq$ 1, must be  satisfied in them.
From Fig.~\ref{3-D}-Bottom  results, that this is true 
only if  $B\leq 10^{8.2}$ G.  
As a consequence, the value of $B \sim 10^9$ G  (given in the literature) 
may be applied only to  Z-type sources with no  radio jet.
Indeed, our upper limit well agrees
with the estimate of Titarchuk et al. (2001)
 on the microquasar Scorpius X-1.
These authors deriving   $B$ from magnetoacoustic oscillations in kHz QPO in 
 three neutron stars (Scorpius X-1 one of them) determine 
a strength of $10^{7-8}$ G on the surface of the neutron star.

\subsection{Millisecond  accretion-powered pulsars} \label{millisec}

Millisecond  accretion-powered pulsars also have a weak magnetic field  $B \sim 10^8 $G together with the main   characteristic of a rapidly spinning neutron star.
They are very few and all of them in the class of LMXRBs.
As one can derive from the values given in Table 1, the millisecond X-ray pulsars
are extreme Atoll sources.
The prolonged, sustained accretion of matter
on the neutron star from the long-living 
companion, carrying angular momentum,
is thought to be responsible for the acceleration to milliseconds.
Less clear is the cause for the decay of B
(Cumming et al. 2001; Chakrabarty 2005; Psaltis 2004).

As shown in Fig. 1 
the obstacle for  jet production in millisecond accretion-powered pulsars
is their low accretion rate.
For the  average  $B\sim 10^8$G , assumed in the literature,
the condition $R_A/R_*<1$ would  be fulfilled only for   
accretion rates of  $\geq 6\times 10^{-9} \rm{M}_{\odot}~\rm{yr}^{-1}$
whereas  the maximum observed accretion rate nearly is one order of magnitude lower, i.e  
 $\dot M \leq 0.7 \times 10^{-9} \rm{M}_{\odot}~\rm{yr}^{-1}$ (Table 2).
On the contrary, if  $\dot M = 0.7 \times 10^{-9}\rm{M}_{\odot}~\rm{yr}^{-1}$,
 the condition $R_A/R_*<1$ would be fulfilled
for $B=10^{7.5}$ G, a value compatible with observations (Table 1).

In  the  accreting millisecond X-ray pulsar
\object{SAX J1808.4-3658}, 
the long-term mean mass transfer rate is $\dot M \simeq 1 \times   10^{-11}\rm{M}_{\odot}~\rm{yr}^{-1}$
(Chakrabarty \& Morgan 1998).
During bright states 
 peak values
of  $\dot M \sim 0.3-0.7 \times 10^{-9}\rm{M}_{\odot}~\rm{yr}^{-1}$
(Chakrabarty \& Morgan 1998; Gilfanov et al. 1998) have been measured and the   
upper limit on the
magnetic field strength is  $B \simlt$\textit{a  few times} $10^7$ G (Gilfanov et al. 1998).
As a matter of fact, in this source  hints for  a radio jet  have been found.
It is one of the  two accreting millisecond X-ray pulsars,
\object{SAX J1808.4-3658} (Gaensler et al.  1999) and IGR J00291+5934 (Pooley 2004),
that have shown transient radio emission related to X-ray outbursts.
Especially interesting is the fact that  the
size of the radio emitting region of \object{SAX J1808.4-3658} 
is   much larger than the separation of the binary system, which is
what would be expected  for expanding  material  ejected from the system
(Gaensler et al. 1999; Migliari $\&$ Fender 2006).
In other words, \object{SAX J1808.4-3658}, which normally behaves like a pulsar,
 could switch to a  microquasar state at  maximum accretion rate.
While  future high resolution radio observations can probe or rule out the presence of
a radio jet, at the moment 
 theory and observations  give positive indications for one.

\section {Discussion and conclusions} 

The results of our  analysis are:
\begin{enumerate}
\item
It is excluded that X-ray accreting-pulsars or in general
neutron stars with strong (i.e. $B\sim 10^{12}$G) magnetic fields
may be associated with a jet, even if accreting at the Eddington critical rate. 
\item
It is known that Z-sources,  ``low" magnetic field neutron stars accreting at the Eddington critical rate, 
 may develop a jet.
In this paper we quantify the magnetic field strength to  $B \simlt 10^{8.2}$.
\item
It is not ruled out that  a millisecond pulsar could  develop a jet, at least for those sources
where  B$ \simlt 10^{7.5}$ G. In this case
the millisecond  pulsar, could
 switch to a microquasar phase during maximum accretion rate.
The millisecond source \object{SAX J1808.4-3658} with such a low $B$,
shows hints for  a radio jet.
\end{enumerate}

\noindent
One of the major open issues concerning millisecond pulsars is the absence
(and possible non-existence) of sub-millisecond pulsars. The spin distribution  sharply
cuts off well before the breakup spin rate for  neutron stars. The physics setting that limit
is unclear (Chakrabarty 2005).
If the jet hypothesis will be proved, than the jet 
might be the suitable agent of angular momentum sink,
as  in the bipolar outflows from young stellar objects.
The  transport rate of angular momentum by the jet can be two thirds or more of the estimated rate transported through the relevant portion of the disk (Woitas et al. 2005).

\begin{acknowledgements}
We gratefully acknowledge the support by the WE-Heraeus foundation.
\end{acknowledgements}

 
              \clearpage

\end{document}